\documentclass[twocolumn,prd,showpacs,superscriptaddress,groupedaddress,preprintnumbers,nofootinbib]{revtex4-1}
\pdfoutput=1
\usepackage{graphicx,amsfonts,amsmath,amssymb,epsfig,color, mathrsfs,latexsym,url,wasysym,float,xfrac,nicefrac}
\newcommand{\be}{\begin{equation}}
\newcommand{\ee}{\end{equation}}
\newcommand{\nn}{\nonumber}
\newcommand{\ba}{\begin{eqnarray}}
\newcommand{\ea}{\end{eqnarray}}
\begin{document}

\title{Split Supersymmetry Breaking from St\"{u}ckelberg Mixing of Multiple $U(1)$'s}

\author{Ghazaal Ghaffari}
\author{Mahdi Torabian$^*$}
\affiliation{Department of Physics, Sharif University of Technology, Azadi Ave, 11155-9161, Tehran, Iran} 

\begin{abstract}
We show that multiple Abelian sectors with St\"{u}ckelberg mass-mixing simply break supersymmetry via Fayet-Iliopoulos $D$-terms and straightforwardly mediate it to the other sectors. This mechanism naturally realizes a split supersymmetry spectrum for soft parameters. 
Scalar squared-masses (holomorphic and non-holomorphic) are induced through sizable portals and are not suppressed. Gaugino masses, $a$-terms and a $\mu$-like term are generated by higher-dimensional operators and are suppressed. The hypercharge is mixed with extra $U(1)$'s, it's $D$-term in non-vanishing and supersymmetry is broken in the visible sector too. Scalar tachyonic directions are removed by unsuppressed interactions and hypercharge is preserved as supersymmetry is broken. Moreover, if a singlet chiral field is charged under additional $U(1)$'s proportional to its hypercharge, new interaction terms in the K\"{a}hler potential and the superpotential are added through St\"{u}ckelberg compensation. In this case supersymmetry is broken via $F$-terms or mixed $F$ and $D$-terms.
\end{abstract}

\preprint{SUT/Physics-nnn}
\maketitle

\subsection*{Introduction} 
Additional Abelian gauge symmetries has long been considered as a natural extension of the Standard Model (SM) or its supersymmetric version. They are predicted from (and vastly ubiquitous in) string compactifications/D-brane constructions and grand unified theories \cite{Leike:1998wr,Rizzo:2006nw,Langacker:2008yv}. Sectors accommodating Abelian multiplets interact through efficient portals. The Hypercharge of the SM mixes with the extra $U(1)$'s through kinetic terms \cite{Holdom:1985ag} and mass terms a la the Stueckelberg mechanism \cite{Kors:2004dx}. Besides rich phenomenological implications for the SM, the physics in the hidden/dark sectors can be probed via these unsuppressed mixings \cite{Dienes:1996zr,Abel:2008ai,Arvanitaki:2009hb,Kors:2004ri,Kuzmin:2003gh, 
Kors:2005uz,Feldman:2006wd,Kuzmin:2002nt,Feng:2014eja,Feng:2014cla}  

On the other hand in a supersymmetric theory, an important task is to find a mechanism for  supersymmetry (SUSY) breaking and computing soft parameters. In phenomenologically acceptable models, SUSY is broken in a secluded sector with no direct coupling to the supersymmetric SM. A mediation mechanism is necessarily devised the nature of which determines soft parameters and thereby the low energy phenomenology (see \cite{Chung:2003fi}). 

In this paper, applying multiple $U(1)$'s with St\"{u}ckelberg mixing, we provide a simple model for SUSY breaking via Fayet-Iliopoulos (FI) $D$-terms \cite{Fayet:1974jb}
. In the simplest realization, there are two $U(1)$ vector multiplets with one St\"{u}ckelberg chiral superfield so that one diagonal vector field is the massless hypercharge. With one non-zero FI parameter, SUSY is generically broken at tree-level and its effect is straightforwardly transferred to the other multiplets. Soft SUSY breaking parameters can be computed. As we will seem, the scalar squared-masses (both holomorphic and non-holomorphic) are directly coupled to the SUSY breaking sectors and therefore are not suppressed. On the other hand, gaugino masses are induced through higher-dimensional operators with are suppressed by the scale of some $R$-symmetry breaking dynamics. This SUSY breaking dynamics is a simple realization of split SUSY  \cite{ArkaniHamed:2004fb,Giudice:2004tc,ArkaniHamed:2004yi}. Moreover, as the hypercharge is mixed with the massive vector, it's $D$-term is non-zero and SUSY is also broken in the SM sector. However as scalars are coupled to the extra $U(1)$ via an efficient portal, the supertrace constraint is alleviated and the mass spectrum is  phenomenologically acceptable. For the same reason, non of scalars are tachyonic as it normally happens in FI SUSY breaking models. In fact, there are universal contributions to the charged scalar fields which can take over charge-dependent sources. Therefore, the gauge symmetry remains intact as SUSY is broken. The scale of SUSY breaking is set by the FI parameter which, for the sake of collider phenomenology, can be dynamically generated in a supersymmetric gauge sector a la retrofitted FI models \cite{Dine:2006gm}.

In a less minimal model, if the (non-Abelian singlet) matter fields are charged under extra $U(1)$'s proportional to their hypercharge (right-handed leptons as candidates) then the St\"{u}ckelberg field can be applied to make new singlets. Consequently, new terms in the K\"{a}hler potential and the superpotential can be assumed. In this case, SUSY can be broken via non-zero $F$-term too.

The structure of this paper is as follows; In the next section we present the general picture of the SUSY breaking dynamics and induced soft parameters. Then we study two simple models to explicitly realize this mechanism. Finally, we summarize in the last section.

\subsection*{SUSY Breaking and Mediation: Generalities} 
To start, we consider $n_V$ Abelian sectors with vector superfields $V^a$ and $n_S$ chiral superfields $S^m$ in non-linear representation of the gauge groups. Besides gauge-invariant kinetic mixing ${\rm f}^{ab}{\rm W}^a\cdot {\rm W}^b$, where ${\rm W}_\alpha^a = -\frac{1}{4}\bar D^2D_\alpha V^a$, Abelian sectors can mix via $n_S$ portals in the K\"{a}hler function as 
\be\label{Kahler-Stuckelberg} K_{\rm Stuckelberg}= M^2\big[2\alpha^a V^a+\beta^m(S^m+S^{m\dagger})M^{-1}\big]^2,\ee
where $a=1,\cdots,n_V$, $m=1,\cdots,n_S$ and $\alpha$, $\beta$ are some constant. It is invariant under the gauge transformations
\ba 
V^{(a)}&\rightarrow& V^{(a)}+\Lambda^{(a)}+\Lambda^{\dagger(a)},\cr \beta^mS^m &\rightarrow& \beta^mS^m-2M\alpha^{a}\Lambda^{a}.\ea

In general, we could use dimensionless gauge singlet  
\be\label{O} {\cal O}=2\alpha^a V^a+\beta^m(S^m+S^{m\dagger})M^{-1},\ee
to parametrize the Kahler potential
\ba\label{Kahler-O} K \supset&& M^2{\cal F}[{\cal O}], \ea
where $\cal F$ is an arbitrary function ${\cal F}\sim{\cal O}+{\cal O}^2+\cdots$ fixed by a specific UV physics. 
We note that the St\"{u}ckelberg K\"{a}hler potential in \eqref{Kahler-Stuckelberg} is $\sfrac{1}{2}M^2{\cal O}^2{\cal F}''[0]$
. The Lagrangian is computed by full superspace integration 
\ba {\cal L} &=& \sfrac{1}{16} M^2{\textstyle\sum}_{n=1}^4\sfrac{1}{n!}\ {\cal F}^{(n)}_{|_{\theta=\bar\theta=0}}\ {\cal D}^2\bar{\cal D}^2{\cal O}^n\cr 
&=&\alpha^aD^a{\cal F'}_|\cr
&+&\sfrac{1}{2!}\big[\sfrac{1}{2}(2\alpha^aA^a_\mu+\beta^m\partial_\mu \sigma^m)^2\!\!+\!2\beta^m\alpha^a(\psi^m\lambda^a\!+{\rm h.c.})
\cr&&\quad\ +\sfrac{1}{2}\beta^m\beta^n\partial_\mu s^m\partial_\mu s^n+i\beta^m\beta^n\psi^m\sigma_\mu\partial_\mu\bar\psi^n
\cr&&\quad\ + 2\beta^m\beta^nF^mF^{n*}\big]{\cal F}_|''\cr
&+&\sfrac{1}{3!}\beta^m\beta^n\beta^p(F^m\psi^n\psi^p+{\rm h.c}+A^m\psi^n\sigma\bar\psi^p){\cal F}'''_|\cr
&+&\sfrac{1}{4!}\beta^m\beta^n\beta^p\beta^q\psi^m\psi^n\bar\psi^p\bar\psi^q{\cal F}''''_|.
\ea
(Pseudo)scalars $s$ and $\sigma$ are the real and imaginary parts of the lowest component of $S$ respectively. Powers of $M$ are suppressed throughout this paper and can be restored on dimensional ground. The K\"{a}hler term \eqref{Kahler-O} gives mass to $n_S$ vector fields and Dirac mass to $n_S$ gauginos. 

Moreover, it contributes to the scalar potential
\be {\cal V}\supset -\alpha^aD^a{\cal F'}_|(s) - 2\beta^m\beta^nF^mF^{n*}{\cal F}_|''(s),\ee
where $s=\beta^ms^m$. 
It induces an effective field-dependent FI parameter $\alpha^a{\cal F}'_|(s)$. It implies that SUSY can be spontaneously broken via non-zero D-terms is the Abelian sectors. We consider two pure $U(1)$ gauge theory with at least one genuine FI parameter $\xi^a$.  Given the gauge kinetic terms, the D-terms are computed
\be -D^a=\alpha^a{\cal F}_|'(s)+\xi^a,\qquad a=1,2.\ee
There are two conditions on one field $s$; thus generically $D$-terms cannot be simultaneously vanishing and SUSY is broken.
The FI parameter $\xi$, introduced through $K\supset 2\xi^aV^a$, sets the scale of SUSY breaking. It can be either induced a la retrofitted FI models \cite{Dine:2006gm} or a field-dependent one from dynamics at higher scale. 
One linear combination of moduli is stabilized at $s_0$ which is found by solving ${\cal V}_{,s}=0$ giving either ${\cal F}_|'(s)_{,s}=0$ or  ${\cal F}_|'(s)=-\xi^a\alpha^a(\alpha^a\alpha^a)^{-1}$.

The SUSY breaking effect can be mediated to the supersymmetric SM sector in many different ways. In this framework, there is a straightforward portal via the dimensionless gauge singlet \eqref{O}. We note that $\langle {\cal O}\rangle_D=\alpha^aD^a$ or generally
\be \langle {\cal F[O]}\rangle_D\supset{\cal F}'_|\textstyle{\sum}_a\alpha^aD^a.\ee
Therefore, the real superfield \eqref{O} can be applied in the K\"{a}hler potential to mediate SUSY breaking and induce soft parameter. In passing we note that there possibly are suppressed $F$-term contributions $\beta^m\beta^nF^mF^{n*}{\cal F}''_|M^{-2}$ that we ignore for the rest of analysis.

To be more specific and for later phenomenological applications, we have in mind a supersymmetric model with gauge group $G\times U(1)_{Y'}\times U(1)^n$ where $G$ is some non-Abelian factor. There are $n$ additional Abelian groups which are mixed with $U(1)_{Y'}$ and there might be or not light matter charged under them. We distinguish $U(1)_{Y'}$ to emphasize that there are $Y'$-charged chiral matter (for the rest of the paper, we reserve index $a$ for extra $U(1)$'s). In fact, the hypercharge is a diagonal subgroup of  $U(1)$'s and in order to have at least one massless $U(1)$ at low scale, there are at most $n$ St\"{u}ckelberg superfields. 
As SUSY is broken in the Abelian sectors with charged matter fields, matter directly feel the SUSY breaking effect. 
In the following we compute soft parameters.

\paragraph{Scalar masses}  are induced through the Kahler potential
\be \label{soft-scalar}K\supset c_i(\phi_i^\dagger e^{2g_GV_G+2q^i_Yg_YV_Y}\phi_i){\cal G}[{\cal O}],\ee
which implies
\be m^2_{0,i}\sim q^i_Yg_Y\langle D_Y\rangle + c_i\textstyle{\sum}_a\alpha_a\langle D_a\rangle{\cal G}'_|.\ee
Interestingly, there is a charge-independent contribution which would lift the supertrace constraint. In fact, SUSY breaking at tree-level in a globally supersymmetric theory with canonical K\"{a}hler potential implies \cite{Ferrara:1979wa}
\be {\rm str}M^2 \equiv (-1)^{2s}(2s+1)M^2_s = -g^aD^a{\textstyle\sum} q_i.\ee
It is not favored phenomenologically as it predicts an sfermion lighter than all fermions. Moreover without superpotential mass parameter, some of them might be tachyonic and break gauge symmetries. However in this framework, we find that the supertrace is modified by the above coupling as 
\be\label{supertrace} {\rm str}M^2 = -\big[g_Y\langle D_Y\rangle{\textstyle\sum}q_i + \textstyle{\sum}_a\alpha_a\langle D_a\rangle{\cal G}'_|{\textstyle\sum}c_i\big].\ee
Apparently, there is a charge-independent source so that the weighted sum could be made positive-definite over the full spectrum or any arbitrary multiplet of particles.
Moreover for the same reason, one can get the MSSM hypercharge FI parameter to break supersymmetry at tree-level through non-zero $D$-term. One can fix parameters in the K\"{a}hler potential so that all scalar directions are non-tachyonic and SUSY breaking minimum occurs at zero field values. Normally, tachyonic directions are removed by superpotential mass terms which is not an option in the MSSM, and so, pure hypercharge $D$-term SUSY breaking was not possible in the MSSM without breaking electromagnetic or color symmetry. 

It is interesting to note that, besides the sizable portal we discussed above, the K\"{a}hler potential \eqref{soft-scalar} induces other important portals
\be {\cal L}\supset (\alpha^aA_\mu^a+\partial_\mu \sigma)^2\phi^*\phi+\alpha^a(\phi^*\chi\lambda^a+{\rm h.c.}).\ee
The first term is called the gauge portal and the second term is an example of portalino.

\paragraph{Gaugino masses} can be worked out from this Kahler potential
\be K\supset c_A{\rm tr}({\rm W}^A\cdot {\rm W}^{A}+{\rm h.c.}){\cal I}[{\cal O}]\Lambda_R^{-1},\ee
where $A$ runs over all Abelian and non-Abelian gauge superfields. This operator in suppressed by some mass scale $\Lambda_R$ at which $R$-symmetry is broken.
Majorana gaugino masses are computed as
\be M_{1/2,A}=c_A(\alpha_YD_Y+\textstyle{\sum}_a\alpha_aD_a){\cal I}_|' \Lambda^{-1}_R.\ee
For $n$ Abelian gauginos, there are also supersymmetric Dirac mass from St\"{u}ckelberg mixing that we must take care of when diagonalizing the gaugino mass matrix.

\paragraph{Holomorphic scalar mass and $a$-terms} are induced through
\be K\supset (W_2+{\rm h.c.}){\cal J}[{\cal O}]+(W_3+{\rm h.c.}){\cal J}[{\cal O}]\Lambda^{-1},\ee
where $W_2$ and $W_3$ include bilinear and trilinear terms respectively. 
It implies that
\ba {\cal L}&\supset& (W_{2|}+{\rm h.c.}){\cal J}_|'(\alpha_YD_Y+\textstyle{\sum}_a\alpha_aD_a)\cr&+&
(W_{3|}+{\rm h.c.}){\cal J}_|'(\alpha_YD_Y+\textstyle{\sum}_a\alpha_aD_a)\Lambda^{-1}.\ea
We note that the holomorphic mass parameters ($B_\mu$-like) are not suppressed. However, the trilinear $a$-terms are suppressed by a factor of $\Lambda$ which is set by UV physics.
We note that, the first K\"{a}hler term also generates suppressed masses for chiral fermions (Higgsino-like)
\be m_{1/2}\supset \beta^mF^{m*}{\cal J}_|'M^{-1},\ee
given that $F$-term components of St\"{u}ckelberg superfields are non-zero.
\paragraph{$\mu(like)$-term} Finally, a $\mu$-like can also be induced as a result of SUSY breaking dynamics via
\be K\supset ({\cal D}^2W_2+{\rm h.c.}){\cal K}[{\cal O}]\Lambda^{-1},\ee
which implies
\be \mu = {\cal K}_|'(\alpha_YD_Y+\textstyle{\sum}_a\alpha_aD_a)\Lambda^{-1}.\ee
The above soft parameters would be seen as a realization of split SUSY spectrum \cite{ArkaniHamed:2004fb,Giudice:2004tc,ArkaniHamed:2004yi}.

\subsection*{Simple Prototype Models}
In this section, we take simple models to study the dynamics of SUSY breaking and compute, in particular, the scalar mass spectrum. Models can be distinguished based on representation of chiral matter under $G\times U(1)_{Y'}\times U(1)^n$. In general, matter fields are in some representation $[{\bf r},q_{Y'},q_a]$ and we assume there is no gauge anomaly. 

We start with a model whose matter fields $\phi_\pm^i$ are in $[{\bf r}({\bf r}'),\pm,0]$  in representation and for now $\bf r\neq r'$. They carry charges of hypercharge and are in some non-trivial representation of $G$  so that no superpotential mass parameter is allowed (as the matter of the MSSM).
The full K\"{a}hler potential is parametrized as 
\be\label{minimal-Kahler} K= \phi_\pm\phi_\pm^\dagger e^{2g_GV_G\pm2g_YV_Y}\big(1+c_\pm{\cal G}[{\cal O}]\big),\ee
where $\cal G$ is an arbitrary function of \eqref{O}.
The hypercharge is mixed with extra $U(1)$'s thus we need to redefine the hypercharge vector superfield so that kinetic terms and the interactions take the canonical form. The $D$-terms are computed from the K\"{a}hler potentials \eqref{Kahler-O} and \eqref{minimal-Kahler} and the gauge kinetic terms
\ba \label{d-term-condition}
-D^a&=&\alpha^a {\cal F}_|'(s)+\alpha^a{\cal G}_|'(s)\big(c_+|\phi_+|^2+c_-|\phi_-|^2\big)+\xi^a,\cr
-D_Y&=&\alpha_Y {\cal F}_|'(s)+g_Y\big(|\phi_+|^2-|\phi_-|^2\big)+\xi_Y.\ea
Generically, there are $n_F$ (for $n_F$ flavor of matter) holomorphic conditions from the superpotential as $\partial_iW[\phi_i]=0$. They can be satisfied if $\phi_i=0$ in the minima.
On the other hand, the $D$-terms cannot be simultaneously vanishing if there are at least {\it one} extra $U(1)$ besides the hypercharge and at least {\it one}  FI parameter. Consequently, supersymmetry is generically broken at tree level in Abelian sectors a la traditional FI mechanism.

The scalar mass of chiral matter are computed
\ba \label{scalar-L} -{\cal L}\supset \big[&\pm& g_Y(\alpha_Y{\cal F}_|'(s)+\xi_Y)\cr&+&c_\pm\alpha^a{\cal G}_|'(s)(\alpha^a{\cal F}_|'(s)+\xi^a)\big]|\phi_\pm|^2.\ea
The first line is the charge-dependent contribution from the hypercharge sector and the second line gives a charge-independent contribution. If the latter is positive-definite and takes over the former, then SUSY is spontaneously broken in a minimum where no scalar component receives a vev. Apparently, there is a wide range of parameters/field values that the above condition is satisfied. 
Thus,  gauge symmetry remains unbroken alongside SUSY breaking.

The SUSY breaking vacuum can be found around  $\langle\phi_i\rangle=0$ using ${\cal V}_{,s}=0$. For a simple model in which ${\cal F=O}^2$ and ${\cal G}={\cal O}$  with one non-zero $\xi_X$ we find
\be {\cal F}_|'(s_0)=-\alpha_X\xi_X(\alpha^A\alpha^A)^{-1}.\ee
Then, the $D$-terms are
\ba \langle D_X\rangle &=& \xi_X\alpha_Y^2(\alpha_A\alpha_A)^{-1},\\
\langle D_{Y} \rangle&=&\xi_X\alpha_X\alpha_{Y}(\alpha_A\alpha_A)^{-1}. \ea
The soft scalar masses are computed 
\ba m_{\pm}^2 &=& \pm g_Y\langle D_Y\rangle+c_\pm\alpha^a\langle D^a\rangle\cr
&=&\xi_X\alpha_X\alpha_Y[\pm g_Y+c_\pm\alpha_Y](\alpha_A\alpha_A)^{-1}.\quad\ \ \ea
which can be made positive definite.

Next, we consider chiral fields in either representation $[{\bf 1},\pm ,0]$ or $[{\bf r}(\bar{\bf r}),\pm ,0]$ so that a supersymmetric mass term in the superpotential $W\supset m\phi_+\phi_-$ is allowed (like the Higgs sector of MSSM). Moreover, the following can be added to the K\"{a}hler potential 
\be K\supset (\phi_+\phi_-+{\rm h.c.})\tilde{\cal G}[{\cal O}],\ee
for arbitrary $\tilde{\cal G}$. The D-terms are computed
\ba \label{d-term-condition}
-D^a&=&\alpha^a {\cal F}_|'(s)+\alpha^a{\cal G}_|'(s)\big(c_+|\phi_+|^2+c_-|\phi_-|^2\big)+\xi^a\cr &&\qquad\quad \, + \alpha^a\tilde{\cal G}_|'(s)(\phi_+\phi_-+{\rm h.c.}),\cr
-D_Y&=&\alpha_Y {\cal F}_|'(s)+g_Y\big(|\phi_+|^2-|\phi_-|^2\big)+\xi_Y\cr &&\qquad\quad\ \, +\alpha_Y\tilde {\cal G}_|'(s)(\phi_+\phi_-+{\rm h.c.}).
\ea
Again with $n_F$ holomorphic conditions, 
$D$-terms and $F$-terms cannot be simultaneously vanishing if there is at least {\it one} extra $U(1)$  and at least {\it one} FI parameter. The scalar masses in \eqref{scalar-L} receive a supersymmetric contribution $-{\cal L}\supset m^2|\phi_\pm|^2$.

In passing we note that matter can be charged under some extra $U(1)$'s. However for the above conclusion to hold, there must be at least one additional $U(1)$ under which matter is neutral {\it i.e.}  in representation $[{\bf r},\pm ,\pm q_1,\cdots,0]$. In general, the scalar masses are
\ba m_{\pm}^2= m^2&\pm& \textstyle{\sum}_iq_ig_i\langle D_i\rangle+c_\pm{\cal G}_|'(s)\textstyle{\sum}_a\alpha_a\langle D_a\rangle,\ea
 $i$ runs over charged $U(1)$'s and $a$ runs over neutral $U(1)$'s
 
Finally, there is an interesting case that $G$-singlet matter are charged under all $U(1)$'s (which are mixed through St\"{a}ckelberg) such that $q_Yg_Y\alpha_Y^{-1} = q_ag_a\alpha_a^{-1}\equiv 2\gamma$. Namely, they are in representation $[{\bf 1},\pm q_Y,\pm q_Yg_Y\alpha_a/g_a\alpha_Y]$. Interestingly,  the SM right-handed leptons can be put in these representations. All $U(1)$ charges are proportional to the hypercharge and thus these representations are anomaly free. The St\"{u}ckelberg superfield, as a compensator, can be applied to make new invariants
\be \phi e^{-\gamma S}\quad {\rm and}\quad \phi^\dagger e^{-\gamma S^\dagger}.\ee
The general  superpotential is
\ba W\sim\phi_i\cdots \phi_je^{-(\gamma_i+\cdots+\gamma_j)S}&\supset& \phi_i e^{-\gamma_i S}+\phi_i^2 e^{-2\gamma_i S}+\cdots,\cr &+&\phi_i\phi_je^{-(\gamma_i+\gamma_j)S}+\cdots.\ea
It is interesting to note that for such representations, linear terms in the superpotential are possible. 
Moreover, the K\"{a}hler potential is extended as follows
\ba K&\sim& (\phi e^{-\gamma S}+{\rm h.c.})\hat{\cal G}[{\cal O}] +\phi\phi^\dagger e^{-\gamma(S+S^\dagger)}\tilde{\cal G}[{\cal O}]\cr &+&\phi\phi^\dagger e^{\pm2g_YV_Y\pm2q_ag_aV_a}{\cal G}[{\cal O}]+\cdots.\ea
New interaction terms are possible through St\"{u}ckelberg compensation. The K\"{a}hler potential is not canonical and thus the K\"{a}hler metric is not diagonal.

To be more explicit, we study a simple model of two chiral multiplets $\phi_\pm$ of charges $(\pm, \pm g_Y\alpha_a/g_a\alpha_Y)$ under $U(1)_Y\times U(1)_a$. 
The leading terms in the superpotential are given by
\ba\label{superpotential-gamma} W&=&
\lambda_\pm\phi_\pm e^{\mp\gamma S} +\mu_\pm \phi_\pm^2 e^{\mp2\gamma S} +\mu\phi_+\phi_-  
\cr &+&\kappa_\pm\phi_\pm^2\phi_\mp e^{\mp\gamma S}+\eta_\pm\phi_\pm^3 e^{\mp3\gamma S}+\cdots.\ea
The leading order terms in the K\"{a}hler potential are
\ba\label{Kahler-gamma} K &\sim& \!\big(\phi_\pm e^{\mp\gamma S}
+ \phi_\pm^2 e^{\mp2\gamma S}+\phi_+\phi_- +{\rm h.c.}\big)\hat{\cal G}[{\cal O}]\cr
&+&\! \big(\phi_\pm\phi_\mp^\dagger e^{\mp\gamma(S-S^\dagger)}\!+{\rm h.c.} + \phi_\pm\phi_\pm^\dagger e^{\mp\gamma(S+S^\dagger)} \big)(1\!+\!\tilde{\cal G}[{\cal O}])\cr
&+&  \phi_\pm\phi_\pm^\dagger e^{\pm2g_YV_Y\pm2q^ag^aV^a}\big(1+{\cal G}[{\cal O}]\big),\ \quad
\ea
where we suppress order one coefficients for brevity. From superpotential we find that
\ba\label{superpotential-gamma} W_{,\pm}&=&
\lambda_\pm e^{\mp\gamma S} \!+ 2\mu_\pm\phi_\pm e^{\mp2\gamma S} \!+ \mu\phi_\mp  
+\cdots,\cr
W_{s} &=& \mp\gamma \phi_\pm e^{\mp\gamma S}+\cdots\ . 
\ea
The $D$-terms in the canonical basis are computed 
\ba -D_a&=&\alpha_a {\cal F}_|'(s)+g_aq_a\big(|\phi_+|^2-|\phi_-|^2\big)+\xi_a\cr &+&\alpha_a\hat {\cal G}_|'(s)(\phi_\pm e^{\mp\gamma s}+\phi_\pm^2 e^{2\mp\gamma s}+\phi_+\phi_-+{\rm h.c.})\cr
&+&\alpha_a\tilde{\cal G}'_|(s)(\phi_\pm\phi_\mp^*e^{\mp\gamma(S-S^*)}+{\rm h.c.}). \ea
Clearly, the $F$ and $D$-terms are not simultaneously vanishing and SUSY is broken. 

Given the general K\"{a}hler potential and superpotential, the scalar potential is computed in the appendix and the full mass spectrum is derived. In particular we find the scalar masses as
\ba m_{\pm}^2= \mu^2+4\mu_\pm^2e^{\mp4\gamma s_0}\pm {\textstyle\sum}_aq_ag_a[\alpha_a{\cal F}'_|(s)+\xi_a].\quad \ea
There are supersymmetric charge-independent contributions to overcome the destabilizing charge-dependent one so that neither fields are tachyonic. Therefore, $\langle\phi_\pm\rangle=0$ is certainly possible for a broad range of parameters. 
Around the above symmetry preserving minimum, the scalar potential is computed
\be {\cal V} =  \sfrac{1}{2}\big[\alpha_a{\cal F}'_|(s)+\xi_a\big]^2+\lambda_{+}^2e^{-\gamma s}+\lambda_{-}^2e^{\gamma s}.\ee
Apparently, the vacuum energy is non-zero and SUSY is  broken at tree-level through mixed F and D-terms  \cite{Fayet:1974jb,ORaifeartaigh:1975nky}. In this case, the FI parameters can be zero or non-zero of either sign. 

\subsection*{Conclusion}
In this paper we presented simple FI SUSY breaking models with multiple Abelian sectors mixed through S\"{u}ckelberg mass-term. SUSY breaking effects are straightforwardly mediated to the SM sector. We observed that this framework predicted a split SUSY breaking spectrum with heavy scalars and light gauginos. At least one FI parameter is needed which sets the scale soft parameters. We also studied a very particular model with matter with hypercharge-mirrored $U(1)$ charges. In that scenario, SUSY is broken via either pure $F$-term or mixed $F$ and $D$-terms. We noted that, although SUSY is also broken in the visible sector, a realistic mass spectrum can be achieved. 
We expect that the MSSM phenomenology in this framework is very rich and we postpone its detailed study to a work in preparation.

\paragraph*{Acknowledgments}
We would like to thanks Fernando Quevedo and Liliana Velasco-Sevilla for discussions. This work is supported by the research office of SUT. We would like to thank IPM for financial support. MT would like to thank ICTP for hospitality during the final stages of this work. 

\ \\$^*$ Electronic address: {\tt mahdi@physics.sharif.edu}

\subsection*{Appendix}
In this appendix, given a general superpotential, K\"{a}hler potential and gauge kinetic term, we compute the mass matrices assuming  St\"{u}ckelberg/kinetic mixing. Generally, the dynamics in the (Abelian) gauge sectors is accounted for through 
\be\label{f-superpotential} {\cal L} \supset \int{\rm d}^2\theta\big({\tt f}^{ab}
+ f_{ab}W\big){\rm W}^a\cdot{\rm W}^b + {\rm h.c.},\ee
where ${\tt f}^{ab}=\frac{1}{8\pi {\rm i}}\Big(\begin{array}{cc} \tau^a  & \epsilon  \\ \epsilon  & \tau^b \end{array} \Big)$ and  $\tau$ is the complexified coupling  $(\theta/2\pi)+{\rm i}(4\pi/g^2)$. Moreover,  we can add the following to the action
\be \label{h-Kahler}{\cal L}\supset \int{\rm d}^2\theta{\rm d}^2\bar\theta h_{ab}({\rm W}^a\cdot {\rm W}^b +{\rm h.c.})K.\ee 
In the above $f$ and $h$ are symmetric matrices in the field space. The kinetic mixing can be removed by diagonalizing $\tt f$, however, the matrices $f$ and $h$ could not be made into diagonal form simultaneously. Gauge dynamics in the non-Abelian sectors follows straightforwardly.

The total action can be read from superspace integration of the above superpotential and the K\"{a}hler potential \cite{Wess:1992cp}. We are interested in SUSY breaking minima and the spectrum of particles around them. Therefore, it is useful to find the mass matrices. We let $I$ runs over chiral superfields $\phi^i$ and $S$ (Note that contributions from \eqref{f-superpotential} and \eqref{h-Kahler} can be taken care of by letting $I$ run over $\phi^i$, $S$ and also ${\rm W}^a\cdot {\rm W}^b$. The scalar, fermion and auxiliary components of the latter includes $-\lambda^a\lambda^b$, $-\lambda^a D^b$ and $D^aD^b$ and we explicitly use in the mass matrices below).

The fermionic mass matrix can be read from fermion bilinear terms  in the total action as follows
\ba
[M_{1/2}]_{IJ} \!&=&\! W_{,IJ} + g_{\bar KI,J} F^{*\bar K}+g_{IJ}h_{ab}D^aD^b,\cr
[M_{1/2}]_{Ia}\! &=& \!-{\rm i}\sqrt 2 {\cal K}_{,I}^a\!-\!\frac{\rm i}{\sqrt 2}f_{ab}W_{\!,I}D^b\!\!+\!2\sqrt 2{\rm i}g_{I\bar J}h_{ab}F^{*\bar J}D^b,\cr
[M_{1/2}]_{ab}\! &=&\! -\frac{1}{2}f_{ab}W_{\!,I}F^I\!\!-\!2h_{ab}K^cD^c\!\! +\!2h_{ab}g_{I\bar J}F^IF^{*\bar J}\!,\ \ \ea
which has both supersymmetric and supersymmetry breaking sources. It is an $(N_f+N_S+N_V)\times(N_f+N_S+N_V)$ matrix which mixes chiral fermion-gaugino spinors in flavor basis.

In the above $g_{I\bar J} = K_{,I\bar J}$ and the Killing potential ${\cal K}^a$ can be determined through \cite{Wess:1992cp}
\ba\label{Killing} g_{I\bar J}{\cal Y}^{*a\bar J} &=& {\rm i}{\cal K}^a_{,I}\quad {\rm and}\quad
g_{I\bar J}{\cal Y}^{aI} = -{\rm i}{\cal K}^a_{,\bar J}.
\ea
The Killing vectors ${\cal Y}^a$ are fixed via gauge transformations $\delta \phi^I = {\cal Y}^{aI} \epsilon^a$. 
It also takes care of \eqref{h-Kahler} if one lets $I$ run over all chiral superfields including ${\rm W}^a\cdot {\rm W}^b$. 
Here, the gauge transformations are as follows
\ba 
\delta s=-\frac{{\rm i}}{2}M \alpha^a \epsilon^a\quad,\quad
\delta \phi^i= -{\rm i} gq_i\phi^i \epsilon,
\ea 

The scalar mass matrix, which receives supersymmetric and non-supersymmetric contributions, is determined from the scalar potential as 
\ba [M_0^2]_{I\bar J} &=& g_{K\bar L, I\bar J}F^KF^{*\bar L} + g^{\bar K\bar L}_{,I}W^*_{,\bar K\bar J}F^{*\bar L}+ g^{KL}_{,\bar J}W_{,KI}F^{L}\cr
&+&g^{K\bar L}W_{,KI}W^*_{,\bar L\bar J} + {\rm f}^{-1ab}{\cal K}_{,I}^a{\cal K}_{,\bar J}^b + {\cal K}_{,I\bar J}^a {\rm D}^a,\cr 
[M_0^2]_{IJ} &=& (g^{KL}W_{,KIJ}+g^{KL}_{,I}W_{,KJ}+g^{KL}_{,J}W_{,KI})F^L\cr  &+& 
g_{K\bar L,IJ}F^KF^{*\bar L} + {\rm f}^{-1ab}{\cal K}^a_{,I}{\cal K}^b_{,J} + {\cal K}^a_{,IJ}{\rm D}^a\cr
&+&f_{ab}D^aD^b W_{,IJ}.
\ea
Finally, the mass matrix of vector fields is given by \cite{Wess:1992cp}
\ba 
[M_1^2]^{ab} &=& 2 g^{I\bar J}{\cal K}^a_{,I}{\cal K}^{b*}_{,\bar J},\ea

The auxiliary fields $F^I$ and $D^a$ can be integrated out from the action by using their equation of motion 
\ba g_{I\bar J}{\rm F}^I -\frac{1}{2}g_{I\bar J,K}\chi^I\chi^K+W^*_{\bar J} +\frac{1}{4} f_{ab}W^*_{\bar J} \bar\lambda^a\bar\lambda^b &=&0,\ \\
 {\rm Re}({\tt f}_{ab}\!\!+\!\!f_{ab}W_0){\rm D}^b \!+\!\frac{1}{2\sqrt{2}}({\rm i}f^{ac}W_{,I}\chi^I\lambda^c\!+\!{\rm h.c.})\!+\!{\cal K}^a\!\! &=&\!0,\quad\ \
\ea
Then, the scalar potential reads as
\ba\label{scalar-potential}
{\cal V} = g^{I\bar J}W_{,I}W^*_{,\bar J} + \frac{1}{2}{\tt f}_{ab}^{-1}{\cal K}^a{\cal K}^b,
\ea
from which we determine the scalar mass matrix
\be M_0^2 = \left(\begin{array}{cc}
{\cal V}_{,I\bar J} & {\cal V}_{,IJ}  \\
{\cal V}_{,\bar I\bar J} & {\cal V}_{,\bar IJ} \end{array}\right).\ee
The fermion mass matrix is written as follows
\ba
[M_{1/2}]_{IJ} &=& W_{,IJ}-\Gamma^K_{IJ}W_{,K}  + g_{I J}{\rm f}^{-1}_{ab}{\rm f}^{-1}_{ac}{\cal K}^b{\cal K}^c,\cr
[M_{1/2}]_{Ia} &=& -{\rm i}\sqrt{2}{\cal K}^a_{,I} \!+\! \frac{{\rm i}}{\sqrt 2}{\cal K}^c{\rm f}^{-1}_{bc}f_{ab,I} \!+\!{\rm i} \sqrt{2}g_{I\bar J}F^{*\bar J}{\rm f}^{-1}_{ab}{\cal K}^b\!,\cr
[M_{1/2}]_{ab} &=&  \frac{1}{2}f_{ab,I}g^{I\bar J}W^*_{,\bar J} + h_{ab}{\rm f}^{-1}_{cd}\alpha^c{\cal K}^d.\ea
Finally using above, the supertrace is computed as  
\ba {\rm str}M^2\!\! &=&2g^{I\bar J}{\cal V}_{,I\bar J} + 6 g^{I\bar J}{\cal K}^a_{,I}{\cal K}^{b*}_{,\bar J}  - 2\delta^{ab}M_{1/2ac}M^\dagger_{1/2cb} \cr &-&\! 2 g^{I\bar J}g^{K\bar L}M_{1/2IK}M^\dagger_{1/2\bar J\bar L}\! -\! 2g^{I\bar J}\delta^{ab}M_{1/2Ia}M^\dagger_{1/2b\bar J}\cr &=& 2{\cal K}{\cal K}_{,I\bar J}g^{I\bar J} - 2 R^{I\bar J}W_{,I}W^*_{,\bar J}-2{\cal K}{\rm f}^{-2}{\cal K}g^{I\bar J}f_{,I}f^*_{,\bar J} \cr  &-&\frac{1}{2}g^{I\bar J}g^{K\bar L}f_{,I}W^*_{,\bar J}f_{,\bar L}^*W^*_{,K} - 2{\cal V}(f_{,I}g^{I\bar J}W^*_{\bar J}+{\rm h.c.}) \cr &-& 4{\rm f}^{-1}{\cal K}g^{I\bar J}({\cal K}_{,I}f^*_{,\bar J}+{\rm h.c.}) -4({\cal K}{\rm f}^{-2}{\cal K})g^{I\bar J}{\rm F}_I{\rm F}_{\bar J}^*\cr &-&2({\cal K}{\rm f}^{-2}{\cal K})(W_{,\bar I\bar I}-\Gamma_{\bar I\bar I}^{\bar J}W_{\bar J}+{\rm h.c.})\cr &+&(N_f+N_S)({\cal K}{\rm f}^{-2}{\cal K})^2+4{\cal K}^a{\rm f}^{-1}_{ab}({\cal K}^b_{,I}F_I+{\rm h.c.})\cr &+&2{\rm f}^{-1}_{bc}{\rm f}^{-1}_{ad}{\cal K}^c{\cal K}^d(f_{ab,I}F_I+{\rm h.c.})- 8{\cal V}^2. \ea

\end{document}